\documentclass{article}

\usepackage{hyperref}
\usepackage{amssymb}
\usepackage{amsmath}
\usepackage{mathtools}
\usepackage{theorem}
\usepackage{tikz}

\usepackage{microtype}

\usetikzlibrary{calc}

\newtheorem{definition}{Definition}
\newtheorem{theorem}{Theorem}

\newcommand{\mb}[1]{{\bf #1}}
\newcommand{\mc}[1]{{\mathcal{#1}}}

\newcommand{\fall}[1]{{\forall\,{#1},\ }}

\DeclareMathOperator\arctanh{arctanh}
\DeclareMathOperator\Ag{Ag}

\renewcommand{\d}{\mathrm{d}}

\makeatletter
\newcommand{\raisemath}[1]{\mathpalette{\raisem@th{#1}}}
\newcommand{\raisem@th}[3]{\raisebox{#1}{$#2#3$}}
\makeatother

\begin{document}

\title{Geometric Time and Causal Time in Relativistic Lagrangian Mechanics}
\author{Olivier Brunet\footnote{\texttt{olivier.brunet} at \texttt{normalesup.org}}}
\maketitle


In this article, we argue that two distinct types of time should be taken into account in relativistic physics: a \emph{geometric} time, which emanates from the structure of spacetime and its metrics, and a \emph{causal} time, indicating the flow from the past to the future. A particularity of causal times is that its values have no intrinsic meaning, as their evolution alone is meaningful. In the context of relativistic Lagrangian mechanics, causal times corresponds to admissible parameterizations of paths, and we show that in order for a langragian to not depend on any particular causal time (as its values have no intrinsic meaning), it has to be homogeneous in its velocity argument.
We illustrate this property with the example of a free particle in a potential. Then, using a geometric Lagrangian (i.e.\ a parameterization independent Lagrangian which is also manifestly covariant), we introduce the notion of ageodesicity of a path which measures to what extent a path is far from being a geodesic, and show how the notion can be used in the twin paradox to differentiate the paths followed by the two twins.

\section{Introduction -- The Two Relativistic Times}

It is usually agreed that the theory of relativity has led one to merge space and time into a single entity, spacetime, in which time and space should be considered on par, the only difference between these two notions being the corresponding sign in the signature of the metrics. But identifying time to a mere dimension of spacetime is, in our opinion, hiding the fact that two types of time do actually intervene and should be taken into account.


The first one is directly related to spacetime and to the corresponding dimension added to the three spatial ones. This kind ot time, which one usually has in mind in a relativistic context, is essentially a ``length'' of a timelike interval divided by the speed of light $c$. To that respect, it is directly related to the geometry of spacetime through its metric, and we call it a \emph{geometric} time. Other typical examples of geometric times are those measured by the proper time along a worldline 
or the time coordinate in a given Lorentz frame (again, divided by~$c$).

However, another important type of time is present in relativity, related to its causal structure and to the flow from the past to the future. This type of time is a way to label the different stages during the temporal evolution of a system. Usual examples of causal times are provided by the values of the indexes labelling the hypersurfaces of a foliation of spacetime. Regarding worldlines, any admissible parameterization provides a causal time, as long as tangent vectors are future-oriented timelike ones. 
But contrary to geometric times, the values of such a time have no intrinsic meaning, and it is the evolution of these values alone which has a meaning as they help distinguish the future from the past. Such times will be called \emph{causal}.

Let us remark that in relativistic Lagrangian mechanics, both types of time are present. First, as spacetime events are, in a given reference frame, determined by for coordinates, one of which corresponding to a geometric time. Thus, in the notation $L(\mb x, \dot {\mb x})$, both $\mb x$ and the 4-vector $\dot{\mb x}$ have a component corresponding to a geometric time. However, the derivation leading to $\dot{\mb x}$ refers to the dynamical evolution of $\mb x$, so that it is made with regards to a causal time
. Similarly, in the Euler-Lagrange equation, the time derivation is also clearly related to a causal time, rather than to a geometric one.

\section{Parameterization Independance}

Having identified two types of time in a relativistic setting, let us focus on causal times and its role w.r.t.\ Lagrangian functions. As expressed previously, the values of causal times do not have any intrinsic meaning. In particular, in relativistic Lagrangian mechanics, as causal times correspond to parameterizations of paths, the fact that they should play no role in the physical behaviour of a system implies that the action along a given path should not depend on the actual admissible parameterization of the path.

In order to determine how this property translates in terms of Lagrangians, consider a path $\mc P$ parameterized as $\mb x(t)$ for $t \in [\alpha, \beta]$. The action of this path is~then
\begin{equation} \label{eq:action1}
S(\mc P) = \int_\alpha^\beta L\bigl(\mb x(t), \dot{\mb x}(t)\bigr) \, \d t
\end{equation}
Any other parameterization of $\mc P$ can be written in the form $\mb y(u)= \mb x \circ \varphi(u)$ with $\alpha = \varphi(a)$, $\beta = \varphi(b)$, and such that $\fall {u \in [a, b]} \varphi'(u) > 0$. Considering this second parameterization, we also want:
\begin{equation} \label{eq:action2}
S(\mc P) = \int_a^b L\bigl(\mb y(u), \dot{\mb y}(u)\bigr) \, \d u
\end{equation}
This expression, in terms of $\mb x$ and $\varphi$, yields
\begin{equation} \label{eq:action3}
S(\mc P) = \int_a^b L\bigl(\mb x \circ \varphi (u), \varphi'(u) \, \dot{\mb x} \circ \varphi (u)\bigr) \, \d u
\end{equation}
But by an elementary change of integration variable (putting $t = \varphi(u)$) in equation~\eqref{eq:action1}, we also obtain
\begin{equation} \label{eq:action4}
S(\mc P) = \int_a^b L\bigl(\mb x \circ \varphi(u), \dot{\mb x} \circ \varphi(u)\bigr) \, \varphi'(u) \, \d u\end{equation}
As we want both equations~\eqref{eq:action3} and~\eqref{eq:action4} to hold for any path and any suitable function $\varphi$, this implies that~$L$ must be such that
$$ \fall {\lambda > 0} L(\mb x, \lambda\, \dot{\mb x}) = \lambda \, L(\mb x, \dot{\mb x}). $$
In other words, a Lagrangian has to be 1-homogeneous\footnote{We recall that a function $f$ is $k$-homogeneous if for all $x$ and $\lambda > 0$, one has $f(\lambda x) = \lambda^k f(x)$.} in its second argument order to have causal times play no role in the value of the action along a path.

\ 


Obviously, as it is well known, this requirement has dramatic consequences regarding Hamiltonian mechanics. We recall Euler's homogeneous function theorem which states that a function is $k$-homogeneous if and only if, for all $\mb x$,
$$ \mb x \cdot \nabla f(\mb x) = k f(\mb x) $$
If we define the conjugate momentum
$$ p^\mu = \frac {\raisemath{-1pt}{\partial L}}{\partial \dot x_\mu}, $$
then Euler's homogeneous function theorem entails
$ p_\mu \dot x^\mu = L $, so that the Hamiltonian obtained as the Legendre Transform of $L$~is
$$ H = p_\mu \dot x^\mu - L = 0 $$
and, as Euler's theorem states an equivalence, this means that a Lagrangian function is parameter independent (i.e.\ causal time independent) iff the associated Hamiltonian is constantly zero.
Conversely, having a non-zero Hamiltonian means that the Lagrangian function relies on some particular causal time.

A consequence of this is that causal time independence implies that one cannot rely on any Hamiltonian-based method, and should rely on Lagrangians instead. And contrary, for instance, to Goldstein \cite{Goldstein:ClassicalMechanics} which states that ``there does not seem to be any compelling reason why the covariant Lagrangian has to be homogeneous in the first degree'', we do believe that being causal time independent is indeed a compelling reason, as for instance it ensures that the action along a path depends only on the spacetime events constituting it, and not on any particular parameterization of the path.






\section{Geometric Lagrangians}

Another natural requirement for a Lagrangian function in a relativistic setting, aside being causal time independent, is to be manifestly covariant, so that it does not depend on a particular reference frame either. In that case, the value of the Lagrangian would only depend on the geometry of a path, and not on any particular curve having the path as its image and expressed in a particular reference frame.

\begin{definition}[Geometric Lagrangian]
A Lagrangian will be said to be \emph{geometric} if it is both Lorentz-covariant, and~$1$-homogeneous in its second argument.
\end{definition}

First, we can remark that the necessity of having a geometric Lagrangian can be used as a guide for designing Lagrangians. In particular, any linear combination of terms of the form
$$ \sqrt{\dot x_\mu \dot x^\mu}, \ A_\mu(\mb x) \dot x^\mu, \ \frac 1 {\sqrt{\dot x^\mu \dot x_\mu}} B_{\nu \eta}(\mb x) \dot x^\nu \dot x^\eta, \ \frac 1 {\dot x^\mu \dot x_\mu} C_{\nu \eta \kappa}(\mb x) \dot x^\nu \dot x^\eta \dot x^\kappa, \ \hbox{etc.} $$
leads to a geometric Lagrangian. 

Consider, for instance, a free particle. It is possible to find a large variety of Lagrangians for it in the literature. For instance, taken from \cite{Goldstein:ClassicalMechanics,JoseSaletan:ClassicalMechanics,Rindler06:Relativity,HobsonEfstathiouLasenby:GR}, up to a multiplicative scalar constant, it is possible to find:
$$
\dot x_\mu \dot x^\mu \qquad \qquad \sqrt{\dot x_\mu \dot x^\mu} \qquad \qquad \sqrt{1 - \frac {\dot x_i \dot x^i}{c^2}} = \sqrt{1 - \beta^2} $$
where $(\dot x_i)$, with $i$ ranging from $1$ to $3$, represents the 3-speed of a particle in the Lorentz frame ``under consideration''. 
The presence of an additional (time independent) potential energy usually leads to the addition of an extra term of the form $ - U(x, y, z) $, leading to a Lagrangian like
$$ - m \sqrt{1 - \beta^2} - U(x,y,z) $$
However, quoting \cite{JoseSaletan:ClassicalMechanics}, ``this treatment of the relativistic particle exhibits some of the imperfections of the relativistic Lagrangian (and Hamiltonian) formulation of classical dynamical systems. For one thing, it uses the nonrelativistic three-vector velocity and position but uses the relativistic momentum. For another, all of the equations are written in the special coordinate system in which the potential is time independent, and this violates the relativistic principle according to which space and time are to be treated on an equal footing.''

Considering the kinetic term alone, it is clear that $\sqrt{1 - \beta^2}$ is not covariant, and that $\dot x_\mu \dot x^\mu$ is not 1-homogeneous so that the only candidate for a geometric Lagrangian is (with the correct multiplicative constants)
$$ - m c \sqrt{\dot x^\mu \dot x_\mu} $$
even though it might look ``awkward'' \cite{Rindler06:Relativity}.

For potential energy, a term of the form $-U(x, y, z)$ is clearly not suitable for a geometric Lagrangian. It is, in particular, not 1-homogeneous in $\dot {\mb x}$. But it can easily be turned into a suitable geometric form the following way. Let~$\mb e_\mu$ be a vector basis for the Lorentz frame $\mc R$ in which $V$ is defined. As it is an energy, it can be seen as the time-component of a 4-vector. And, indeed, if one defines:
$$ \mb A(c t, x, y, z) = \frac {V(x, y, z)} {\raisemath{1pt}c} \mb e_0, $$
it is then easy to verify that the term $ - A_\mu(\mb x) \, \dot x^\mu $ leads to the correct equation of motion using the geometric Lagrangian
\begin{equation} \label{eq:pot_geo}
- m c \sqrt{\dot x_\mu \dot x^\mu} - A_\mu(\mb x) \, \dot x^\mu = 
- m c \sqrt{\dot {\mb x} \cdot \dot {\mb x}} - \frac {V(\mb x)} {\raisemath{1pt}c} \, \mb e_0 \cdot \dot {\mb x}
\end{equation}




Geometric Lagrangians are also very closely related to the Euler-Lagrange equation. Consider again the equality
$$ L = \frac {\partial L}{\partial \dot {\mb x}} \cdot \dot {\mb x} $$
verified by a geometric Lagrangian. If we differentiate this expression w.r.t.\  parameter~$t$
, we~get
\begin{multline*}
\frac {\partial L}{\partial \mb x} \cdot \frac {\d \mb x}{\d t} + \frac {\partial L}{\partial \dot {\mb x}} \cdot \frac {\d \dot {\mb x}}{\d t} = \Bigl(\frac {\d}{\d t} \frac {\partial L}{\partial \dot {\mb x}} \Bigr) \cdot \dot {\mb x} + \frac {\partial L}{\partial \dot {\mb x}} \cdot \frac {\d \dot {\mb x}}{\d t} \\
\iff \frac {\partial L}{\partial \mb x} \cdot \dot {\mb x} + \frac {\partial L}{\partial \dot {\mb x}} \cdot \ddot {\mb x} = \Bigl(\frac {\d}{\d t} \frac {\partial L}{\partial \dot {\mb x}} \Bigr) \cdot \dot {\mb x} + \frac {\partial L}{\partial \dot {\mb x}} \cdot \ddot {\mb x} \\
\iff \frac {\partial L}{\partial \mb x} \cdot \dot {\mb x} - \Bigl(\frac {\d}{\d t} \frac {\partial L}{\partial \dot {\mb x}} \Bigr) \cdot \dot {\mb x} = 0 \\
\iff \Bigl( \frac {\partial L}{\partial \mb x} - \frac {\d}{\d t} \frac {\partial L}{\partial \dot {\mb x}} \Bigr) \cdot \dot {\mb x} = 0
\end{multline*}
In the final equality, we recognize 
$$ \frac {\partial L}{\partial \mb x} - \frac {\d}{\d t} \frac {\partial L}{\partial \dot {\mb x}} $$
which appears in the Euler-Lagrange equation.
As $L$ is covariant, this expression actually corresponds to a~$4$-vector, and the previous 
equality shows that it is indeed orthogonal to $\mb {\dot x}$. In particular, it is spacelike and, as $\mb {\dot x} \neq \mb 0$, we have
$$ \frac {\partial L}{\partial \mb x} - \frac {\d}{\d t} \frac {\partial L}{\partial \dot {\mb x}} = \mb 0 \iff \Bigl\|\frac {\partial L}{\partial \mb x} - \frac {\d}{\d t} \frac {\partial L}{\partial \dot {\mb x}}\Bigr\| = 0 $$
where we define the norm of a vector $\mb v$ as $\bigl\|\mb v\bigr\| = \sqrt{\bigl|v_\mu v^\mu\bigr|}$. This suggests the following definition:

\begin{definition}[Ageodesicity]
The \emph{ageodesicity} of a path $\mc P$ parameterized as $\bigl\{ \mb x(t) \bigm| t \in [a,b] \bigr\}$ is the real number
$$ \Ag(\mc P) = \int_{\mc P} \Bigl\|\frac {\partial L}{\partial \mb x} - \frac {\d}{\d t} \frac {\partial L}{\partial \dot {\mb x}}\Bigr\|\, \d t$$
\end{definition}

Let us first prove that the ageodesicity of a path is a purely geometric quantity, i.e.\ it is covariant and does not depend on the parameterization of the path. Indeed, with the previous notations, remembering that $\varphi'(t) > 0$ and since $\partial_2 L$ is $0$-homogeneous in its second argument, i.e.\ $\partial_2 L(\mb x, \lambda \dot {\mb x}) = \partial_2 L(\mb x, \dot {\mb x})$\footnote{From here on, $\partial_1 L$ corresponds to $\frac {\partial L}{\partial \mb x}$, and $\partial_2 L$ to $\frac {\partial L}{\partial \dot {\mb x}}$.} (as follows from the 1-homogeneity of $L$ in its second argument), we~have
\begin{multline*}
\partial_1 L\bigl(\mb y(u), \dot{\mb y}(u)\bigr) = \partial_1 L\bigl(\mb x \circ \varphi (u), \varphi'(u) \, \dot{\mb x} \circ \varphi (u)\bigr) \\ = \varphi'(u) \, \partial_1 L\bigl(\mb x \circ \varphi (u), \dot{\mb x} \circ \varphi (u)\bigr)
\end{multline*}
and
\begin{multline*}
\bigl[u \mapsto \partial_2 L\bigl(\mb y(u), \dot{\mb y}(u)\bigr)\bigr]'(u) = \bigl[u \mapsto \partial_2 L\bigl(\mb x \circ \varphi (u), \varphi'(u) \,\dot{\mb x} \circ \varphi (u)\bigr)\bigr]'(u) \\
= \bigl[u \mapsto \partial_2 L\bigl(\mb x \circ \varphi (u), \dot{\mb x} \circ \varphi (u)\bigr)\bigr]'(u) = \bigl[\bigl(t \mapsto \partial_2 L(\mb x(t), \dot {\mb x}(t)) \bigr) \circ \varphi \bigr]'(u) \\
= \varphi'(u) \bigl[\bigl(t \mapsto \partial_2 L(\mb x(t), \dot {\mb x}(t)) \bigr)\bigr]'\bigl(\varphi(u)\bigr)
\end{multline*}
so that, considering the change of variable $t = \varphi(u)$,
\begin{multline*}
\Bigl\| \partial_1 L\bigl(\mb y(u), \dot{\mb y}(u)\bigr) - \bigl[u \mapsto \partial_2 L\bigl(\mb y(u), \dot{\mb y}(u)\bigr)\bigr]'(u) \Bigr\| \, \d u \\
= \Bigl\| \varphi'(u) \, \partial_1 L\bigl(\mb x \circ \varphi (u), \dot{\mb x} \circ \varphi (u)\bigr) - \varphi'(u) \, \bigl[t \mapsto \partial_2 L\bigl(\mb x (t), \dot{\mb x} (t)\bigr)\bigr]'\bigl(\varphi(u)\bigr) \Bigr\| \, \d u \\
= \Bigl\| \partial_1 L\bigl(\mb x \circ \varphi (u), \dot{\mb x} \circ \varphi (u)\bigr) - \bigl[t \mapsto \partial_2 L\bigl(\mb x (t), \dot{\mb x} (t)\bigr)\bigr]'\bigl(\varphi(u)\bigr) \Bigr\| \, \varphi'(u) \, \d u \\
= \Bigl\| \partial_1 L\bigl(\mb x(t), \dot{\mb x}(t)\bigr) - \bigl[t \mapsto \partial_2 L\bigl(\mb x (t), \dot{\mb x} (t)\bigr)\bigr]'(t) \Bigr\| \, \d t 
\end{multline*}
With reasonable assumptions of continuity and of smoothness, a path has its ageodesicity equal to $0$ iff the vector $\partial_1 L - \frac {\d}{\d t} \partial_2 L$ is null all along, i.e.\ iff it is indeed a geodesic:

\begin{theorem}
A path $\mc P$ is a geodesic w.r.t.\ a geometric Lagrangian if and only if it verifies $$\Ag(\mc P) = 0$$
\end{theorem}

\noindent More generally, the ageodesicity of a path is a geometric measure of how far it is from being a geodesic. In the next section, we will present an application of this measure.

\section{Ageodesicity for a Free Particle, and the Twin Paradox}
Let us consider again the geometric Lagrangian of a free particle (without potential energy):
$$ L = - m c \sqrt{\dot {\mb x} \cdot \dot {\mb x}} $$
In this situation, the Euler-Lagrange vector is simply
$$ - \frac {\d}{\d t} \frac {m c \dot {\mb x}}{\sqrt{\dot{\mb x} \cdot \dot{\mb x}}} $$
so that
$$ \Ag(\mc P) = \int_{\mc P} \Bigl\| \frac {\d}{\d t} \frac {m c \dot {\mb x}}{\sqrt{\dot{\mb x} \cdot \dot{\mb x}}} \Bigr\|\, \d t $$





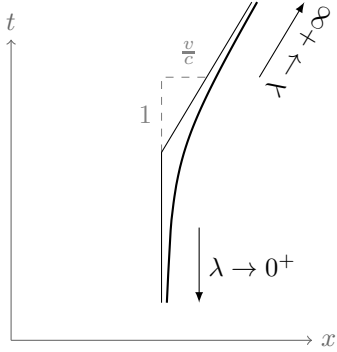
\begin{figure}
\begin{centering}
\begin{tikzpicture}
\draw [thin, black!60,->] (-2,-2.5) -- (2,-2.5) node [right] {$x$};\draw [thin, black!60,->] (-2,-2.5) -- (-2,1.5) node [above] {$t$};

\draw [thin, dashed, black!50] (0,0)-- (0,1) node [midway, left] {$1$}  -- (0.6,1)  node [midway, above] {$\ \frac v {\raisemath{1.5pt}c}$}  ;
\draw [thin] (0, -2) -- (0, 0) -- (1.2, 2);
\draw [thick, domain=0.236068:4.23607,smooth,variable=\t] plot({3*\t/10},{(\t-1/\t)/2}) ;
\draw [-latex] (0.5,-1) -- (0.5, -2) node [midway, right] {$\lambda \rightarrow 0^+$};
\draw [-latex] (1.3,1) -- (1.9, 2) node [midway, below, sloped] {$\lambda \rightarrow +\infty$};
\end{tikzpicture} \\
\end{centering}
\caption{Change of direction} \label{fig:twin}
\end{figure}
\begin{figure}
\begin{centering}
\begin{tikzpicture}
\draw [domain=-2:2,smooth,variable=\t,black!10] plot({3/10*(\t+sqrt(\t*\t+0.0390625))},{\t}) ;
\draw [domain=-2:2,smooth,variable=\t,black!20] plot({3/10*(\t+sqrt(\t*\t+0.078125))},{\t}) ;
\draw [domain=-2:2,smooth,variable=\t,black!30] plot({3/10*(\t+sqrt(\t*\t+0.15625))},{\t}) ;
\draw [domain=-2:2,smooth,variable=\t,black!40] plot({3/10*(\t+sqrt(\t*\t+0.3125))},{\t}) ;
\draw [domain=-2:2,smooth,variable=\t,black!50] plot({3/10*(\t+sqrt(\t*\t+0.625))},{\t}) ;
\draw [domain=-2:2,smooth,variable=\t,black!60] plot({3/10*(\t+sqrt(\t*\t+1.25))},{\t}) ;
\draw [domain=-2:2,smooth,variable=\t,black!60] plot({3/10*(\t+sqrt(\t*\t+2.5))},{\t}) ;
\draw [domain=-2:2,smooth,variable=\t,black!60] plot({3/10*(\t+sqrt(\t*\t+5))},{\t}) ;
\draw [domain=-2:2,smooth,variable=\t,black!60] plot({3/10*(\t+sqrt(\t*\t+10))},{\t}) ;
\draw [domain=-2:2,smooth,variable=\t,black!60] plot({3/10*(\t+sqrt(\t*\t+20))},{\t}) ;
\draw [domain=-2:2,smooth,variable=\t,black!60] plot({3/10*(\t+sqrt(\t*\t+40))},{\t}) ;
\draw [thick] (0, -2) -- (0, 0) -- (1.2, 2);
\draw [-latex] (2.08,-1) -- (0.3, -1) node [midway, below, sloped] {$\beta \rightarrow 0$};
\end{tikzpicture} \\
\end{centering}
\caption{Change of direction, variation of $\beta$} \label{fig:twinbis}
\end{figure}
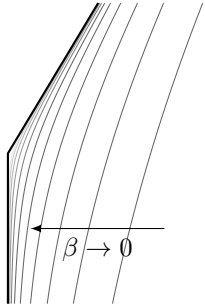

Let us compute the exact ageodesicity of a path corresponding to a change of velocity. We consider the family of paths, as represented on figure~\ref{fig:twin},
defined for all $t \in \mb R^{\star +}$~by
$$ X^\mu(t) = \beta \bigl(c(t - \frac \alpha t), v t, 0, 0\bigr) $$
where $\alpha$ and $\beta$ are parameters. In this case, one~has
$$ \Bigl\| \frac {\d}{\d t} \frac {m c \dot{\mb x}}{\sqrt{\dot{\mb x} \cdot \dot {\mb x}}} \Bigr\| = \frac {2 m v \alpha t c^2}{(\alpha + t^2)^2 c^2 - v^2 t^4}, $$
one primitive of which~being
$$ 
\frac { m c} 2 \ln \Bigl(\frac{\alpha c + (c + v) \lambda^2}{\alpha c + (c - v) \lambda^2}\Bigr) 
$$
Integrating from $0$ to $+\infty$, this leads~to
$$ \Ag(\mc P) = \frac { m c} 2 \ln \frac {c + v}{c - v} =  m c \arctanh \frac v {\raisemath{1pt}c}$$
Up to the factor $ m c$, we recognize the asymptotic change of rapidity.
It can also be remarked that this result neither depends on $\alpha$ nor on $\beta$, the two parameters we had introduced for the sake of generality. And as, when $\beta$ tends to $0$, the limit path corresponds to an instantaneous change of velocity of $v$ as shown in figure~\ref{fig:twinbis}, this means that such a change of velocity does indeed lead to an increase of ageodesicity of
$$  m c \arctanh \frac v {\raisemath{1pt}c} $$



As an application of this result, let us consider the ``twin paradox''. 
We recall that this paradox involves two twins, one of whom makes a journey into space in a high-speed rocket and returns home to find that his twin, who has remained on Earth, has aged more. The puzzling aspect of this result is that there seems to be a symmetry between the two twins, each of them seeing the other as moving w.r.t.\ himself. However, this interpretation is rather  naive, as one twin remains in a single inertial frame while the trajectory of the other twin, the one in the rocket, involves two frames: one for the outbound journey and another for the inbound one.

In terms of ageodesicity, the two trajectories can easily be distinguished: the twin who remains on Earth follows a geodesic (if we neglect the movements she makes on Earth), so that the ageodesicity of its path is approximately $0$. Meanwhile, if the other twin starts its journey by travelling away from Earth at speed $v$, and then travels back at the same speed, the change of direction entails an ageodesicity~of
\begin{equation} \label{eq:twi_age}
 m c \arctanh \Bigl(\frac {2 v c}{c^2 + v^2}\Bigr)
\end{equation}
The total ageodesicity associated to the travelling twin is even greater, as we have neglected the ageodesicity contributions corresponding to the initial acceleration and the final deceleration.

\begin{figure}
\begin{centering}
\begin{tikzpicture}
\node (B) at (1.6, 0) {} ;
\draw [black!50] (B) circle (6pt) ;
\draw [thick] (0, -2) -- (0, 2) node [pos = 0.5] (A) {};
\draw [thick] (0, -2) -- (1.6, 0) node (B) {} -- (0, 2) ;
\draw [stealth-] (A) -- ($(A) + (-0.5cm, 0)$) node [left] {$\Ag = 0$} ;
\draw [fill] (0, -2) circle (1pt) ;
\draw [fill] (0, 2) circle (1pt) ;
\draw [stealth-] ($(B) + (6pt, 0)$) -- ($(B) + (6pt + 0.5cm, 0)$) node [right] {$\Ag = m c \arctanh \Bigl(\dfrac {2vc}{c^2 + v^2}\Bigr)$} ;
\end{tikzpicture} \\
\end{centering}
\caption{Twin Paradox} \label{fig:twin2}
\end{figure}
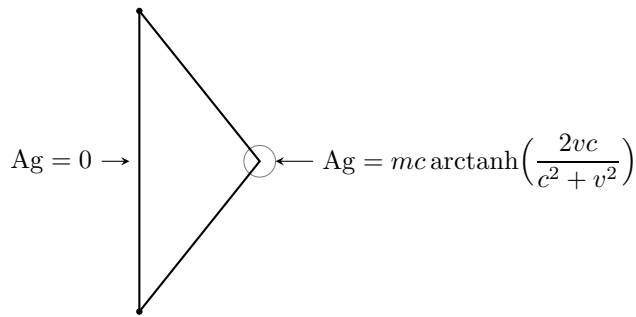

This illustrates the fact that the ageodesicity of a path provides an intrinsic, geometric measure of how far is a path from a geodesic and that, in the context of the twin paradox, it provides a means to distinguish both trajectories. 

\section{Conclusing Remarks}


In this article, we have just initiated the study of causal time in Lagrangian mechanics and introduced the notion of causal time independent Lagrangian and, more generally, of geometric Lagrangians. The latter notion, in particular, shall prove to be especially useful in the context of relativistic Lagrangian mechanics. The fact that any admissible parameterization shall lead to the correct result (should it be an equation of motion, an action along a path or the ageodesicity of a path) is a rather convenient property and 
the situation is, to that respect, similar to the use of covariant expressions which ensures that the correct results will be obtained, whichever reference obtained is used. 

In the example of a free particle, it can be remarked that the parameterization of the path was such that the parameter could hardly be seen as a geometric time. However, it is an admissible causal time and thus, leads to the correct actions and ageodesicities, as the Lagrangian is geometric.

In the case of multiple particles, the improvement of the situation is even more evident. In order to have a covariant formulation, one usually imposes the use of an affine parameter for a single particle, and this requirement cannot be extended to a system of $n$ particles, as it is usually not possible to define a common affine parameter. The usual solution is, instead, to consider a preferred frame, at the cost of losing the generality of the result. But again, if the considered Lagrangian function is geometric, one is assured to obtain general and covariant results.

Finally, quoting Dirac \cite{Dirac33:Lagrangian}, ``there are reasons for believing that the Lagrangian [formulation] is the more fundamental [than the Hamitonian one. In particular,] the Lagrangian method can easily be expressed relativistically, on account of the action function being a relativistic invariant; while the Hamiltonian method is essentially non-relativistic in form, since it marks out a particular time variable as the canonical conjugate of the Hamiltonian function.'' We do believe that one can go even further with this statement, as having a non-vanishing Hamiltonien is precisely the sign that the corresponding Lagrangian function is not causal time independent. 

\bibliographystyle{alpha}

\end{document}